\documentstyle[12pt]{article}
\def\be{\begin{equation}}
\def\ee{\end{equation}}
\def\({\left (}
\def\){\right )}

\begin{document}
 \begin{titlepage}
\vspace*{-4ex}
 \null \hfill ITP-SB-94-038\\
 \null \hfill hep-ph/XXXXX\\
 \null \hfill July, 1994\\
\vskip 1.7 true cm
 \begin{center}
 {\bf \LARGE On Destabilizing Divergences in Supergravity
Models\footnote{This work was supported by the
National Science Foundation under grant PHY--93--09888.}}\\[11ex]
{\large
  Vidyut Jain }
   \\ [2ex]
%

{\large \it
Institute for Theoretical Physics, State Univ. of New York,
 \\ Stony Brook, New York 11794-3840}\\ [2ex]
 \end{center}
\vskip 1.9cm
 \begin{abstract}
 We show that, for standard N=1 supergravity coupled to chiral matter, if the
K\"ahler potential has no terms linear in chiral superfields, then none
are generated through quadratically divergent one-loop corrections. This
holds whether or not the K\"ahler potential contains terms cubic in
chiral (and antichiral) matter superfields.
If however the cosmological constant vanishes and $susy$ is spontaneously
broken, linear terms are present and new quadratically divergent linear
terms may be generated due to the presence of nonvanishing K\"ahler
curvature.
These can potentially destabilize the weak-scale hierarchy.
We discuss the importance
of this for effective supergravity models which contain standard model
singlets.
 \end{abstract}

 \end{titlepage}
\newpage
\setcounter{page}{1}
Standard N=1 supergravity
in four dimensions can be considered as the "leading approximation"
of an effective low energy theory that incorporates the effects of
integrating out physics above the Planck scale. Such a low energy theory
will in general contain an infinite number of nonrenormalizable operators,
and its Lagrangian will contain an arbitrary number of space-time
derivatives. Standard supergravity on the other hand, although
nonrenormalizable, contains at most two space-time derivatives and
furthermore the couplings of gravity and chiral matter are described by
a single function, the K\"ahler potential $G$:
\be G=K + \ln W +\ln \bar{W} \ee
where $K$ is a real function of chiral and antichiral superfields and
$W$, the superpotential, is a holomorphic function of the chiral superfields
$\Phi^i,$ $i=1...N_S$.
In the following we shall use $G$, $K$, and $W$ to denote both superfield
quantities $G(\Phi,\bar{\Phi})$, $K(\Phi,\bar{\Phi})$ and $W(\Phi)$ and
also their scalar components $G(\phi,\bar{\phi})$, $K(\phi,\bar{\phi})$
and $W(\phi)$, respectively. We will however distinguish between chiral
superfields and their scalar components by case (e.g. $\phi=\Phi|$,
$m=M|$).

 The bosonic part of the supergravity Lagrangian is [1]
\be
  {1\over \sqrt{g}} L_B = G_{i\bar{j}} D_\mu\phi^i g^{\mu\nu}
     D_\nu\phi^{\bar{j}}-e^{G}[G^{i\bar{j}}G_i G_{\bar{j}}-3]
     +{1\over2} R,\ee
Here, we use the notation
$\overline{\phi^i}=\phi^{\bar{i}}$.
The K\"ahler metric is given by
\be
    G_{i\bar{j}}=K_{i\bar{j}}={\partial^2 G\over \partial\phi^i
        \partial\phi^{\bar{j}} },\ee
its inverse is $G^{i\bar{j}}=
K^{i\bar{j}}$, $G^{i\bar{j}}G_{k\bar{j}}=\delta^i_k$,
and $G_i=\partial_iG={\partial\over\partial\phi^i} G$.
Throughout, $K_{i\bar{j}}$ ($K^{i\bar{j}}$) is used to lower (raise)
indices.
Finally, the derivatives $D_\mu$ are general coordinate covariant.

In order to successfully describe the world we observe, the matter
content of the supergravity model must include standard model (SM)
fields. In addition, there may be various SM singlets, some of which may
carry non-standard-model gauge group charges. Typically, one expects
some of these SM singlets to develop $vev$s at some scale below $M_{pl}$
and spontaneously break supersymmetry.
Then, supersymmetry breaking terms
are generated for light superfields and
in order to have acceptable masses
for the standard model particles, we typically require
$<e^G>=m^2_{3/2}\sim m^2_W$.
However, the SM singlet $vev$(s) responsible for this may lie anywhere
between $M_{pl}$ and $m_W$.

In superstring models, SM singlets which develop $vev$s include the
dilaton and moduli fields [2,3]. Furthermore, if the gauge group contains a
$U(1)$ factor whose renormalizable couplings give triangle anomalies which
are cancelled by the Green-Schwarz mechanism [4], then some of the fields
charged under this $U(1)$ can develop $vev$s [5].

In these models one generally believes it is non-perturbative physics in
a "hidden" sector which is largely responsible for $susy$ breaking.
The hidden
gauge and chiral fields (which interact only gravitationally with
observable sector fields) are charged w.r.t. a gauge group that becomes
strongly interacting at some scale $\Lambda_{c} < M_{pl}$. At this
scale one expects hidden sector gauginos to form condensates [6,7];
integrating out hidden sector fields gives rise to an effective theory for the
observable sector whose K\"ahler potential includes moduli and dilaton
dependent terms that spontaneously break supersymmetry [2,3,8].

Here, we are not interested in details of the $susy$ breaking terms in
$G$ which are in any case not well understood. Instead we
will assume that the effective theory below some scale $\Lambda_{eff}$ can be
described by a K\"ahler potential of the form
\be K=\tilde{K}(M,\bar{M})+Z_{I\bar{J}}Q^I Q^{\bar{J}}
    +{1\over2}(H_{IJ}Q^I Q^J+h.c.)+...
\ee
\be W=\tilde{W}(M,\bar{M})+{1\over2}\mu_{IJ}Q^I Q^J +
       Y_{IJK} Q^I Q^J Q^K + ...
\ee
Here $M^a$ parameterize chiral superfields whose scalar components
acquire large $(>>m_W)$ $vev$s and $Q^I$ parameterize the remaining
fields (by assumption there are no terms linear in $Q$).
The quantities $Z$, $H$, $\mu$ and $Y$ are in general $M$, $\bar{M}$
dependent.
The ellipsis indicates terms higher order in $Q$,$\bar{Q}$.

The K\"ahler potential describes the effective tree theory of the
fields $M$ and $Q$ below some scale $\Lambda_{eff}$. We have
$m_W^2=<e^{\tilde{K}}|\tilde{W}|^2>$; for the fields $Q$ this leads
to supersymmetric masses [9,10] of $O(e^{\tilde{K}/2}<\mu_{IJ}>+
m_W<H_{IJ}>)$, nonsupersymmetric masses $m^2_{I\bar{J}}$ of
$O(m_W^2 <Z_{I\bar{J}}>)$ and nonsupersymmetric masses $m^2_{IJ}$
of $O(m_W^2<H_{IJ}>+m_W e^{\tilde{K}/2}<\mu_{IJ}>)$.

Thus, the SM $\mu$ term, $W\ni \mu H_1 H_2$, can potentially
receive contributions from both $<\mu_{IJ}>$ and $<H_{IJ}>$; the
contribution from $<H_{IJ}>$ is not naturally too large once
$<e^G>=m_W^2$. We will assume, for whatever reason, the contribution
of $<\mu_{IJ}>$ to the $\mu$-term is of $O(m_W)$.

The issue we address in this article is whether divergent one-loop
corrections in the effective theory can destabilize the model and lead
to large ($>>m_W$) corrections to low energy masses. Being an effective
theory, the divergences should be cut off at $\Lambda_{eff}$.

In particular, if ${G}$ receives quadratically divergent corrections
linear in gauge singlets which couple to SM Higgs fields then large
Higgs masses ($>>m_W$) may be generated even if they were absent at
tree level. One of the ways in which this may potentially happen is through
nonrenormalizable cubic couplings between Higgses and singlets in ${K}$.
Consider [11]
\be K\ni aN \bar{H}_1 H_1/M_{pl}+bN\bar{H}_2H_2/M_{pl}+h.c. \ee
and
\be W\ni gNH_1 H_2 + m_N N^2 + ...\ee
for some singlet $N$. Then, the second term in V, eq. (2), when expanded
about background $vev$s gives a $susy$ breaking contribution
\be V\ni -3 {m_W^2\over M_{pl}} (a n h_1^{\dag} h_1+bn h_2^{\dag} h_2+h.c.).\ee
One might then expect a quadratically divergent correction to $V$ from
one--loop tadpoles involving the Higgses [11],
\be
V_{eff}\ni -O\({\Lambda_{eff}^2\over 16\pi^2}\) {m_W^2\over M_{pl}}
   (a+b)n + h.c.\ee
Such a linear term in $n$, which should be interpreted as coming from
a correction to $K$, indicates that $<n>\neq 0$ at one-loop. It is
then straightforward to check that the trilinear coupling of $N$
to the Higgses through the superpotential will give rise to [11]
\be V_{eff}\ni O\({\Lambda_{eff}^2\over 16\pi^2}\)
    {m_W^2\over M_{pl}m_N} g(a+b) h_1 h_2+h.c.\ee
Thus, if such a linear term does arise, we require (for $|a|,|b|,|g|\sim 1$)
\be \Lambda_{eff}\leq O(4\pi\sqrt{M_{pl}m_N}), \ee
in order to keep the SM masses of $O(m_W)$. This would be
an additional phenomenological constraint on the theory.

Such linear corrections to $K$ are well known to happen in globally
supersymmetric models with explicit breaking terms [12]. In terms of
superfields, if
\be S_{break}=\int d^4 x d^4\theta U N\bar{H}_1H_1+h.c.,\ee
where the spurion field $U=const\cdot\theta^2\bar{\theta}^2$, then
\be \delta S \sim \int d^4x d^4\theta UN+h.c.\ee
due to a one--loop superspace tadpole diagram.

However, as we now demonstrate, if ${G}$ contains no terms linear
in chiral superfields then none are generated by quadratically divergent
one-loop corrections. It is sufficient to consider the non-gauge
interactions since $K$ cannot receive perturbative corrections linear in
gauge nonsinglets and gauge singlets have no gauge interactions.

The crucial difference between the spurion example above and the effective
supergravity theory is that while
a cubic term like (6) in $K$ leads to the same type of $susy$ breaking
term in $V$ as (12), it also leads to a renormalization of the kinetic term for
the various fields. Then, since the fields are not canonically
normalized, the loop corrections to $V$ do not just depend on the
potential and, in fact, terms linear in fields cancel out.

To be more precise, we introduce the variable
\be A(\phi)=e^{K(\phi,\bar{\phi})} W(\phi), \ee
for which
   \be A_{\bar{i}}=K_{\bar{i}} A,\quad A_{i\bar{j}}=K_{i\bar{j}}A
        +K_{\bar{j}} A_i. \ee
We use a formalism which retains manifest
holomorphic field redefinition invariance
[13,14,15,16]. Then $A_{ij}=D_i D_j A$ is defined using derivatives covariant
w.r.t. holomorphic field redefinitions. On a vector $V_j$,
$D_i V_j=\partial_i V_j+\Gamma^l_{ij}V_l$, where the K\"ahler connection
is $\Gamma^l_{ij}=K^{l\bar{m}}\partial_i\partial_j\partial_{\bar{m}}K$.
Then we find for the background field dependent mass-squared matrix [17]
\be
V_{i\bar{j}}=e^{-K}\(A_{ki}\bar{A}_{\bar{j}\bar{l}}K^{k\bar{l}}
   -A_i\bar{A}_{\bar{j}}+K_{i\bar{j}}A_k\bar{A}_{\bar{l}}K^{k\bar{l}}
  -2K_{i\bar{j}}A\bar{A}+R^k_{n\bar{j}i}K^{n\bar{l}}A_k\bar{A}_{\bar{l}}\)
. \ee
In this expression the K\"ahler curvature is
\be R^i_{j\bar{m}k} = -\partial_{\bar{m}}\Gamma^i_{jk},
  \quad R^i_{jkl}=0,\quad R_{\bar{n}jk\bar{m}}=-R_{\bar{n}j\bar{m}k},
      \quad etc. \ee

In (16) it is the second to last term, $V_{i\bar{j}}\ni -2e^{-K}K_{i\bar{j}}
A\bar{A},$ which contains the $susy$ breaking term linear in $(N+\bar{N})$
when $K$ includes terms like (6) [just put $A$ to its $vev$ and pick
$i,\bar{j}$ to correspond to the Higgses]. As
the kinetic terms (2) depend on the K\"ahler metric $K_{i\bar{j}}$, it is
well known that the scalar one--loop quadratically divergent corrections
to $V$ are proportional to [17]
\be K^{i\bar{j}}V_{i\bar{j}} =
    e^{-K}\(A_{ik}\bar{A}^{ik}
  +(N_S-1)A_i\bar{A}_{\bar{j}}K^{i\bar{j}}-2N_SA\bar{A}+
     R^k_{n\bar{m}l}K^{\bar{m}l} A_k\bar{A}^n\).\ee
Chiral fermions, with Lagrangian [1,17]
\be {1\over\sqrt{g}} L_F \ni
  iK_{i\bar{m}}
    \(\bar{\chi}^{\bar{m}}_L\gamma_\mu D^\mu\chi_L^i
      +\bar{\chi}^i_R\gamma_\mu D^\mu\chi_R^{\bar{m}}\)
   -e^{-K/2}\(A_{ij}\bar{\chi}^i_R\chi^j_L+h.c.\),\ee
have a mass matrix proportional to $e^{-K/2} A_{ij}$. Then, a one-loop
quadratically divergent correction from chiral fermions gives exactly
a contribution that cancels the first term in
(18) [17]. This is just the usual cancellation between bosons and
fermions with supersymmetric mass terms.

Thus, the quadratically divergent corrections to $V$ are of three types:
\be V_{eff}\ni {\Lambda_{eff}^2} V,\quad
    \Lambda_{eff}^2 e^G,\quad \Lambda_{eff}^2 R^k_{n\bar{m}l}
      K^{l\bar{m}}A_k\bar{A}_{\bar{l}}K^{n\bar{l}}.\ee
[We have rewritten some terms using the expression for $V$, eq. (2).]
It is straightforward to verify that for $G$ given by (4),(5), neither
$V$ nor $e^G$ contain terms linear in fields $Q^I$. We now focus our
attention on the last expression in (20). If we consider
only fields $Q$ then this expression gives no linear terms in $Q^I$
because $A_I=\partial_I A$ is already at least linear in $Q$. For terms
that involve $M^a$, we examine this expression when the scalar components
$m^a$ of $M^a$ are expanded about their $vev$s: $m^a=<m^a>+\hat{m}^a,
A=<A>+<A_a>\hat{m}^a+<A_{\bar{a}}>\hat{m}^{\bar{a}}+...$. If $A$
contains no terms linear in $\hat{m}^a$, i.e. $<A_a>=0$ for all $a$, then
the last expression in (20) vanishes whenever either $k$ or $\bar{l}$
correspond to fields $\hat{m}$, i.e. there are no terms linear in
$Q^I, Q^{\bar{I}}$ when the fields $m$ are set to their $vev$s.

The conditions $<A_a>=0$, although ensuring $<V_a>=0$, are incompatible
with vanishing cosmological constant and broken $susy$. $<V>=0$
implies
\be <A_a \bar{A}_{\bar{b}} K^{a\bar{b}}>=3<A\bar{A}>\neq 0. \ee
If we require $<V>=0$ then $<A_a>\sim O(m_W M_{pl})$ for some $a$.
Then, the last term  in (20) gives rise to terms linear in $Q$. To
see this, we compute
\begin{eqnarray}
<\partial_I(R^{k\bar{n}j}_{\;\;\;\;\; j}A_k\bar{A}_{\bar{n}})>
   &=&<(\partial_I R^{a\bar{b}j}_{\;\;\;\; j})A_a\bar{A}_{\bar{b}}>
    +<R^{a\bar{N}j}_{\;\;\;\;\;\; j}A_a\partial_I\bar{A}_{\bar{N}}>
   \nonumber \\ & &
+<R^{N\bar{a}j}_{\;\;\;\;\;\; j}\bar{A}_{\bar{a}}\partial_I A_N>
\end{eqnarray}
at $Q=0$. If $K$ contains terms cubic in fields $Q^I,Q^{\bar{I}}$ then
$<\partial_I R^{a\bar{b}j}_{\;\;\;\; j}>$,
$<R^{a\bar{N}j}_{\;\;\;\;\;\; j}>$
and $<R^{N\bar{a}j}_{\;\;\;\;\;\; j}>$ are nonzero. We have
\begin{eqnarray}
<R^{a\bar{N}j}_{\;\;\;\;\;\; j}>&=&
 <K^{\bar{L}J}K^{\bar{N}R}K^{a\bar{b}}K^{\bar{M}P}(\partial_{\bar{b}}
    \partial_P\partial_{\bar{L}}K)(\partial_{\bar{M}}\partial_R
    \partial_J K)>
     \nonumber \\ & & -<K^{\bar{L}J} K^{\bar{N}R} K^{a\bar{b}}
    \partial_{\bar{b}}\partial_R\partial_J\partial_{\bar{L}} K >
     \nonumber \\ & =&
{\overline{<R^{N\bar{a}j}_{\;\;\;\;\;\; j}>}}
\end{eqnarray}
The expression for $<\partial_I R^{a\bar{b}j}_{\;\;\;\;\;\; j}>$
is not very illuminating; however one can check that it
is at least of $O(\partial_Q\partial_Q\partial_{\bar{Q}} K)$
or $O(\partial_{\bar{Q}}\partial_{\bar{Q}}\partial_Q K)$.

Thus $K$ must contain
couplings $f(M,\bar{M})QQ\bar{Q}+h.c.$ for new linear terms to be generated.
With $<A_a>\sim O(m_W M_{pl})$
and $K^{a\bar{b}}\partial_{\bar{b}}K\sim
O(K/M_{pl})$ we (very crudely) estimate
\be V_{eff}\ni O({\Lambda_{eff}^2 m_W^2 \over M_{pl} }) K^{I\bar{J}}
   ( \partial_N\partial_I\partial_{\bar{J}} K) q^N + h.c. \ee
For $K$ containing terms like (6)
this includes corrections like (9). We stress, however, that the origin
of such linear terms is due to nonrenormalizable
couplings in
$K$ $and$ nonvanishing $<A_a>$; fields $Q$ by themselves do not
give rise to linear terms as they do in a globally supersymmetric
model with explicit breaking terms like (12).

As mentioned, such terms can lead to a large SM $\mu$-term
unless the singlets are very massive or $\Lambda_{eff}$ is sufficiently
small. One generally expects cubic couplings in $K$ unless their is
a symmetry reason for their suppression.
In string models we have to consider fields which are SM singlets but are
nonetheless not singlets under the full gauge group. Gauge invariance
forbids terms linear in these fields; however, if part of the gauge group
is spontaneously broken then linear terms may be generated through
higher order terms. Some of the fields charged w.r.t. the broken
group may get large tree-level $vev$s ($>>m_W$) and belong to the set
$M^a$; the rest belong to the set $Q^I$. Let us donate two representative
examples of such fields $\chi$ ($\in M^a$) and $\tilde{\chi}$
($\in Q^I$). For a phenomenologically viable model the couplings of
$\chi$ to SM nonsinglets must be such that $\mu_{IJ}(<\chi>)\leq
O(m_W)$. Even assuming this is arranged, large SM masses may be
generated due to radiatively generated linear terms in $\tilde{\chi}$
and couplings like $W\ni \tilde{\chi}H_1 H_2$.

Of particular interest is the case of an "anomalous" $U(1)_A$ factor
whose field theoretic triangle anomalies are cancelled by the Green-Schwarz
mechanism [4]. In these models, we expect that SM singlet fields $\chi$
which get $vev$s with $<D>=0$ have $U(1)_A$ charges with the same
sign, otherwise $<W>$ will generically be too large.
[There is some combination of these fields $\chi$ and the dilaton
whose imaginary part is "eaten" by the $U(1)_A$ gauge boson.]
 Then  fields
$\tilde{\chi}$ which carry opposite sign charges can typically
have large ($>>m_W$)  masses through superpotential couplings to
fields $\chi$. However, fields $\tilde{\chi}$ that carry charges with
the same sign as $\chi$ can not get large masses in such a manner and
can potentially give the largest corrections to the SM $\mu$-term through
radiative corrections.  An analysis of whether the dangerous couplings in
$K$ actually occur or not is beyond the scope of this article. However,
it may be possible to assign $U(1)_A$ charges so that superpotential
terms like $W\ni g(<\chi>)\tilde{\chi} H_1 H_2$ are absent or strongly
suppressed [18] for such $\tilde{\chi}$.

We close with two issues. First, we have only analyzed radiative corrections
linear in fields $Q^I$. What about corrections quadratic in $Q$? In
fact, one can check that the first two terms in (20) only correct the
tree level mass parameters by a factor of $O(\Lambda_{eff}^2/M_{pl}^2)$.
The last term in (20) can potentially generate mass terms that were absent
at tree level; however any mass term which comes from (a renormalized)
$K$ is not naturally too large.

Finally, we discuss corrections due to loops involving gravitons/gravitinos,
which must necessarily be included in order to get the complete quadratically
divergent corrections. For example, it is obvious that graviton
loops will give
corrections like $\Lambda_{eff}^2 V$ since the bosonic Lagrangian
$L_B\ni -\sqrt{g}V$ contains a background field dependent "mass" term
for the gravitons [17] which is linear in $V$. A careful analysis of the
one--loop corrections to $V_{eff}$ requires gauge fixing of the
graviton and gravitino sectors, including corrections from the
corresponding ghosts and ghostinos. The complete (gauge fixing dependent)
divergent one--loop corrections for supergravity+chiral matter have
been computed in [17].
It was found that all quadratically divergent corrections are of the
form (20). Although $V_{eff}$ is gauge fixing dependent, physical
parameters should not be. Therefore, the additional corrections do not
modify the analysis presented here.

{\bf Acknowledgements.}
I am very grateful to Robert Shrock for the lengthy discussions
which led me to investigate the problem of radiative destabilization.

{\bf References.} \\
{\footnotesize
{[1]}
E. Cremmer, S. Ferrara, L. Girardello, B. Julia, P. van
Nieuwenhuizen and J. Scherk, Nucl. Phys. {\bf B147:} 105 (1979);
E. Cremmer, S.Ferrara, L. Girardello, A. van Proeyen,
Nucl. Phys. {\bf B212:} 413 (1983) \\
{[2]} M. Dine, R. Rohm, N. Seiberg and E. Witten, Phys. Lett. {\bf 156B:}
55 (1985) \\
{[3]} See for example: P. Bin\'etruy and M. K. Gaillard, Nucl. Phys.
{\bf B358:} 121 (1991) \\
{[4]} M. B. Green and J. H. Schwarz, Phys. Lett. {\bf 149B:} 117 (1984) \\
{[5]} M. Dine, N. Seiberg and E. Witten, Nucl. Phys. {\bf B289:} 589 (1987)\\
{[6]} H. P. Nilles, Phys. Lett. {\bf B115:} 193 (1982) \\
{[7]} S. Ferrara, L. Girardello and H.P. Nilles, Phys. Lett.
{\bf B125:} 457 (1983)\\
{[8]} G. Veneziano and S. Yankielowicz, Phys. Lett. {\bf B113}: 231 (1982)\\
{[9]} S.K. Soni and H. A Weldon, Phys. Lett. {\bf B126:} 215 (1983)\\
{[10]} V. Kaplunovsky and J. Louis, Phys. Lett. {\bf B306}: 269 (1993)\\
{[11]} J. Bagger and E. Poppitz, Phys. Rev. Lett. {\bf 71:} 2380 (1993)\\
{[12]} S.J Gates, M. Grisaru, M. Rocek and W. Siegel, Superspace
(Benjamin/Cummings, $\quad$ Menlo Park, CA, 1983)\\
{[13]} S. Honerkamp, Nucl. Phys. {\bf 36:} 130 (1973)\\
{[14]} L. Alvarez-Gaum\'e, D.Z. Freedman and S. Mukhi,
Ann. Phys. (N.Y.) {\bf 134:} 85 (1981) \\
{[15]} S. Mukhi, Nucl. Phys. {\bf B264:} 640 (1986)\\
{[16]} M. K. Gaillard, Nucl. Phys. {\bf B268:} 669 (1986)\\
{[17]} M.K. Gaillard and V. Jain, Phys. Rev. {\bf D49:} 1951 (1994)\\
{[18]} V. Jain and R. Shrock, work in progress
}


\end{document}